\documentclass[twoside,conference,twocolumn,final,12pt,english]{IEEEtran}
\usepackage{pslatex}
\usepackage[T1]{fontenc}
\usepackage[latin1]{inputenc}
\usepackage{graphicx}
\usepackage{babel}

\makeatletter

\providecommand{\tabularnewline}{\\}

\makeatother
\begin{document}

\title{Improved Noise Weighting in CELP Coding of Speech - Applying the Vorbis Psychoacoustic Model To Speex}
 
\author{Jean-Marc Valin$^{12}$, 
         Christopher Montgomery$^{23}$ \\
$^1$ CSIRO ICT Centre, PO Box 76, Epping, NSW, 1710, Australia \\
$^2$ Red Hat, 10 Technology Park Drive, Westford, MA, 01886, USA \\
$^3$ Xiph.Org Foundation, http://www.xiph.org/ \\
Correspondence should be addressed to jmvalin@jmvalin.ca}

\maketitle
\begin{abstract}
One key aspect of the CELP algorithm is that it shapes the coding noise using a simple, yet effective, weighting filter. In this paper, we improve the noise shaping of CELP using a more modern psychoacoustic model. This has the significant advantage of improving the quality of an existing codec without the need to change the bit-stream. More specifically, we improve the Speex CELP codec by using the psychoacoustic model used in the Vorbis audio codec. The results show a significant increase in quality, especially at high bit-rates, where the improvement is equivalent to a 20\% reduction in bit-rate. The technique itself is not specific to Speex and could be applied to other CELP codecs.
\end{abstract}

\section{Introduction}

When the Code-Excited Linear Prediction (CELP) \cite{Schroeder1985}
scheme was originally proposed, one key aspect of the algorithm was
how the noise was shaped using a simple, yet effective, weighting
filter. Since then, audio coding advances have provided significantly
better psychoacoustic models for shaping coding noise. However, psychoacoustic
modeling in CELP has remained essentially the same. In this paper,
we propose to improve the quality of an existing CELP codec by using
better noise shaping, without modifying the bit-stream of the codec.
More specifically, we improve the Speex%
\footnote{http://www.speex.org/%
} CELP codec by using a psychoacoustic model derived from the Vorbis%
\footnote{http://www.vorbis.org/%
} audio codec.

Speex is an open-source multi-rate CELP codec supporting both narrowband
and wideband speech. Unlike most current CELP codecs, it uses a 3-tap
pitch predictor and sub-vector innovation quantization. Vorbis is
a high-quality open-source audio codec designed for music and uses
the modified discrete cosine transform (MDCT). Both Speex and Vorbis
are developed within the Xiph.Org Foundation.

This paper is organized as follows. Section \ref{sec:The-Speex-Codec}
introduces the Speex codec used in this work. Section \ref{sec:The-Vorbis-Psycoacoustic}
then describes the psychoacoustic model used by the Vorbis codec.
The application of that model to Speex is described in Section \ref{sec:Improved-Psycho-acoustics-in-CELP}.
Results are presented in Section \ref{sec:Evaluation-and-Results}
with a discussion in Section \ref{sec:Discussion-and-Conclusion}.

\section{The Speex Codec}

\label{sec:The-Speex-Codec}

Speex is an open-source codec based on the Code-Excited Linear Prediction
(CELP) algorithm. It is targeted mainly towards voice over IP (VoIP)
applications so it is designed to be robust to lost packets. Speex
supports multiple bit-rate, ranging from 2.15 kbps to 24.6 kbps in
narrowband (8 kHz) operation and from 3.95 kbps to 42.2 kbps in wideband
(16 kHz) operation. Some additional features in Speex are:

\begin{itemize}
\item Embedded wideband coding
\item Variable bit-rate (source controlled)
\item Voice activity detection (VAD) and discontinuous transmission (DTX) 
\item Variable search complexity
\end{itemize}
The Speex bit-stream was frozen in March 2003 with the release of
version 1.0. However, since there is no {}``bit-exact'' specification,
it is still possible to improve the quality of the encoder as long
as no modification is required on the decoder side.

\subsection{Perceptual Weighting}

In order to maximize speech quality, CELP codecs minimize the mean
square of the error (noise) in the perceptually weighted domain. This
means that a perceptual noise weighting filter $W(z)$ is applied
to the error signal in the encoder. In most CELP codecs, $W(z)$ is
a pole-zero weighting filter derived from the linear prediction coefficients
(LPC), generally using bandwidth expansion. Let the spectral envelope
be represented by the synthesis filter $1/A(z)$, CELP codecs typically
derive the noise weighting filter as: \begin{equation}
W(z)=\frac{A(z/\gamma_{1})}{A(z/\gamma_{2})}\label{eq:gamma-weighting}\end{equation}
where $\gamma_{1}=0.9$ and $\gamma_{2}=0.6$ in the Speex reference
implementation.

\begin{figure}
\begin{center}\includegraphics[width=1\columnwidth,keepaspectratio]{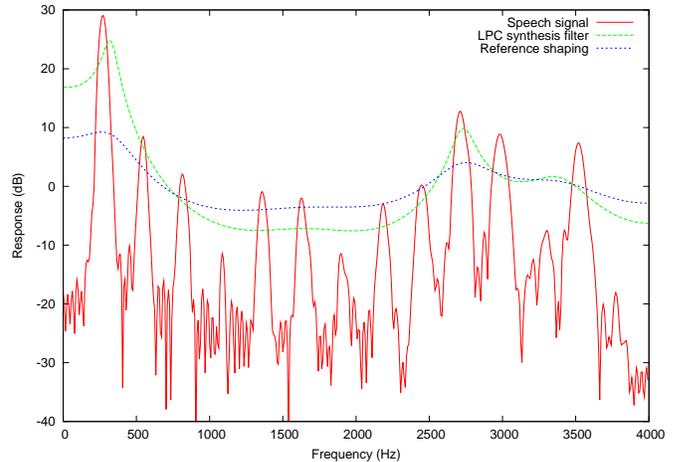}\end{center}

\caption{Standard noise shaping in CELP. Arbitrary y-axis offset.\label{cap:Standard-noise-shaping}}
\end{figure}

The weighting filter is applied to the error signal used to optimize
the codebook search through analysis-by-synthesis (AbS). This results
in a spectral shape of the noise that tends towards $1/W(z)$. While
the simplicity of the model has been an important reason for the success
of CELP, it remains that $W(z)$ is a very rough approximation for
the perceptually optimal noise weighting function. Fig. \ref{cap:Standard-noise-shaping}
illustrates the noise shaping that results from Eq. \ref{eq:gamma-weighting}.
Throughout this paper, we refer to $W(z)$ as the noise weighting
filter and to $1/W(z)$ as the noise shaping filter (or curve).

\subsection{Narrowband Encoder Structure}

\begin{figure}
\begin{center}\includegraphics[width=0.9\columnwidth,keepaspectratio]{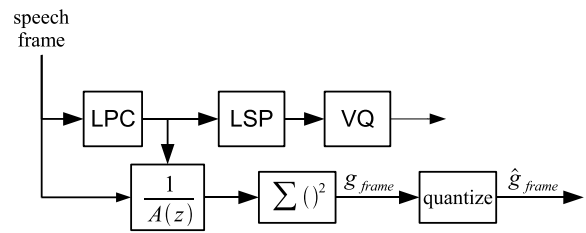}\end{center}

\caption{Frame open-loop analysis\label{cap:Frame-open-loop-analysis}}
\end{figure}

In narrowband, Speex frames are 20 ms long (160 samples) and are subdivided
in 4 sub-frames of 5 ms each (40 samples). For most narrowband bit-rates
(8 kbps and above), the only parameters encoded at the frame level
are the Line Spectral Pairs (LSP) and a global excitation gain $g_{frame}$,
as shown in Fig. \ref{cap:Frame-open-loop-analysis}. All other parameters
are encoded at the sub-frame level.

Linear prediction analysis is performed once per frame using an asymmetric
Hamming window centered on the fourth sub-frame. The linear prediction
coefficients (LPC) are converted to line spectral pairs (LSP) and
vector-quantized using 30 or 18 bits (depending on the bit-rate used).
To make Speex more robust to packet loss, no prediction is applied
on the LSP coefficients prior to quantization. For each sub-frame,
the LSP coefficients are interpolated linearly based on the current
and past quantized LSP coefficients and converted back to the LPC
filter $\hat{A}(z)$. The non-quantized interpolated filter is denoted
$A(z)$. 

\begin{figure}[t]
\begin{center}\includegraphics[width=0.85\columnwidth,keepaspectratio]{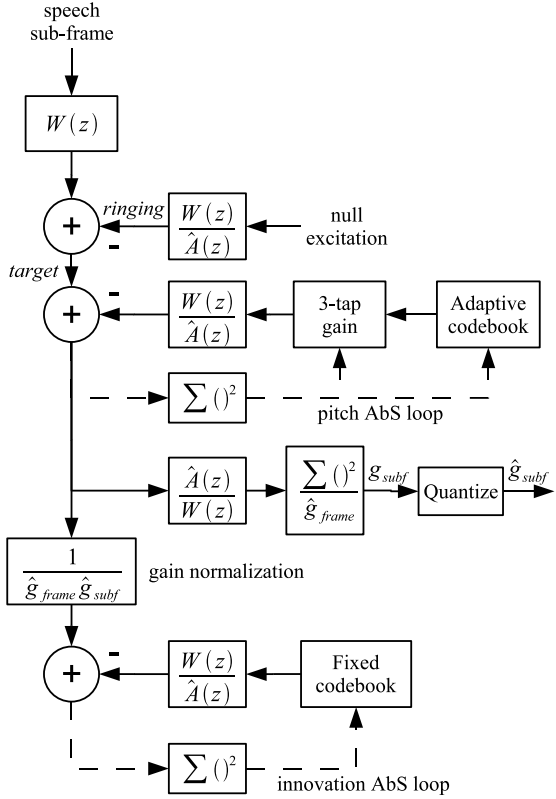}\end{center}

\caption{Analysis-by-synthesis closed-loop optimization on a sub-frame.\label{cap:Sub-frame-AbS}}
\end{figure}

The analysis-by-synthesis (AbS) encoder loop is described in Fig.
\ref{cap:Sub-frame-AbS}. There are three main aspects where Speex
significantly differs from most other CELP codecs. First, while most
recent CELP codecs make use of fractional pitch estimation \cite{Kroon1990}
with a single gain, Speex uses an integer to encode the pitch period,
but uses a 3-tap predictor \cite{Chen1995} (3 gains). The adaptive
codebook contribution $e_{a}[n]$ can thus be expressed as:\begin{equation}
e_{a}[n]=g_{0}e[n-T-1]+g_{1}e[n-T]+g_{2}e[n-T+1]\label{eq:adaptive-3tap}\end{equation}
where $g_{0}$, $g_{1}$ and $g_{2}$ are the jointly quantized pitch
gains and $e[n]$ is the codec excitation memory. 

Many current CELP codecs use moving average (MA) prediction to encode
the fixed codebook gain. This provides slightly better coding at the
expense of introducing a dependency on previously encoded frames.
A second difference is that Speex encodes the fixed codebook gain
as the product of the global excitation gain $g_{frame}$ with a sub-frame
gain corrections $g_{subf}$. This increases robustness to packet
loss by eliminating the inter-frame dependency. The sub-frame gain
correction is encoded before the fixed codebook is searched (not closed-loop
optimized) and uses between 0 and 3 bits per sub-frame, depending
on the bit-rate.

The third difference is that Speex uses sub-vector quantization of
the innovation (fixed codebook) signal instead of an algebraic codebook.
Each sub-frame is divided into sub-vectors of lengths ranging between
5 and 20 samples. Each sub-vector is chosen from a bitrate-dependent
codebook and all sub-vectors are concatenated to form a sub-frame.
As an example, the 3.95 kbps mode uses a sub-vector size of 20 samples
with 32 entries in the codebook (5 bits). This means that the innovation
is encoded with 10 bits per sub-frame, or 2000 bps. On the other hand,
the 18.2 kbps mode uses a sub-vector size of 5 samples with 256 entries
in the codebook (8 bits), so the innovation uses 64 bits per sub-frame,
or 12800 bps.

\section{The Vorbis Psychoacoustic Model}

\label{sec:The-Vorbis-Psycoacoustic}

The masking model we use in this paper is based on elements of the
psychoacoustic model of the Vorbis open-source audio codec, specifically
noise masking. Noise masking in the Vorbis codec is implemented in
a conceptually similar manner to the Spectral Flatness Measure introduced
by Johnston \cite{Johnston1988}. A geometric median and envelope
follower are constructed by smoothing the log spectrum with a sliding
window of approximately one Bark. The distance between the two curves
provides a tonality estimate for a given band of the spectrum. The
envelope curve is companded according to the distance in order to
directly compute a \emph{noise mask} curve that is used for spectral
weighting.

\subsection{Psychoacoustics Basic Ideas}

Audio coding uses psychoacoustics in two ways. The first is a separable
mechanism that quantifies the relative importance of specific audio
features so that the encoding may be weighted toward representing
the most important. This analytical optimization is entirely encoder-side
and although it operates within the framework of the codec design,
it tends not to be part of a codec specification. The metrics and
decision logic may change without affecting the compatibility of the
bit-stream. In this role, psychoacoustics are essential to efficiency
but optional to correct mechanical operation. Most codecs, including
Vorbis and Speex fall in this category.

%The psychoacoustics can be either explicit, when a resolution is assigned for each time-frequency sample, or implicit when psychoacoustics affects how codewords are selected. Most codecs, including Vorbis (explicit psychoacoustics) and Speex (implicit psychoacoustics) fall in this category. 

%Some audio codec incorporate psychoacoustics in a second way. They rely on some amount of psychoacoustic design hardcoded into the codec specification. Such hardwired psychoacoustics are part of the required mechanisms of both the encoder and decoder. A number of contemporary codecs dispense with the latter 'optional' psychoacoustics altogether, relying entirely on the hardwired psychoacoustics of the codec's explicit design. The transform-coded excitation (TCX) algorithm falls in this category.

Secondly, every audio codec also relies on some amount of psychoacoustic
design hardcoded into the codec specification. Such hardwired psychoacoustics
are part of the required mechanisms of both the encoder and decoder.
A number of contemporary codecs dispense with the former \emph{optional}
psychoacoustics altogether, relying entirely on the hardwired psychoacoustics
of the codec's explicit design. In such codecs, like the transform-coded
excitation (TCX) \cite{Lefebvre1994} algorithm, it is not possible
to change the psychoacoustic model without changing the bit-stream.

\begin{figure}
\includegraphics[width=1\columnwidth,keepaspectratio]{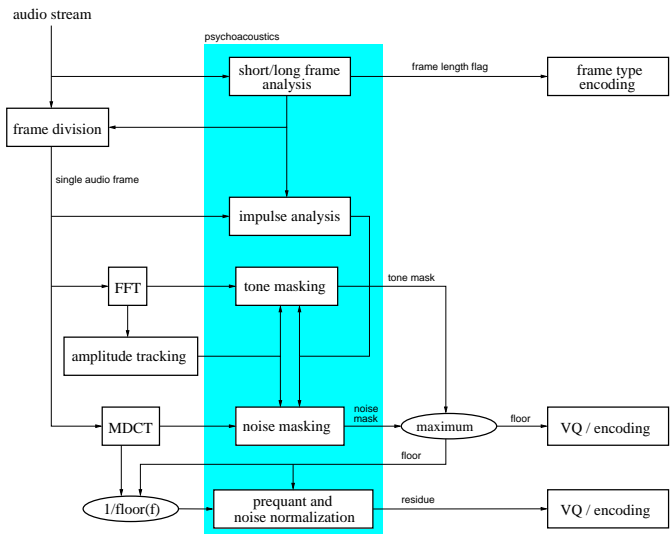}

\caption{Block diagram of the Vorbis encoder applied to a mono input stream.
The shaded block contains the psychoacoustic elements.\label{cap:Block-diagram-Vorbis}}
\end{figure}

\subsection{The Vorbis Model}

\label{sub:The-Vorbis-Model}

The Vorbis codec relies on encode-side psychoacoustic heuristics in
order to produce effective bit-streams, as illustrated in Fig. \ref{cap:Block-diagram-Vorbis}.
 From each frame's audio spectrum, it computes a \emph{floor} curve
which is used as a unit-resolution whitening filter, that is, the
floor is used to normalize the spectrum such that a direct linear
quantization of the normalized spectrum is perceptually appropriate.
Thus, the floor performs a similar function to the conceptually simpler
perceptual weighting filter used in Speex and other CELP codecs. However,
in the case of Vorbis, the floor curve is required for decoding, whereas
CELP \emph{merely} uses the weighting filter for optimizing the search
for the best entries in the codebooks.

The frame-by-frame psychoacoustic metrics used by the Vorbis encoder
to compute the floor fall into four main categories: tone masking,
noise masking, noise normalization, and impulse analysis.

\subsubsection{Tone Masking}

Tone masking in the reference Vorbis encoder is a simple tone-tone
masking engine which produces a single spectral curve per frame. This
curve represents the threshold of perception, per spectral line, for
a pure tone in the current spectral frame. The curve is computed by
superposition of tonal masking curves such as those originally presented
by Ehmer \cite{Ehmer1959}. Note that the reference encoder uses curves
measured independently by Xiph.Org.

Tone-tone masking analysis and the tonal masking curve are of only
marginal direct importance in the modern Vorbis codec; most commercially
produced audio programmes have seen deep single- and multi-band compression
such that there is insufficient dynamic depth for tone masking to
result in much savings.

\subsubsection{Noise Masking}

Noise masking in Vorbis is something of a historical misnomer; although
it began as a simple noise-noise masking mechanism analogous to the
tone-tone masking above, in time it has become more concerned with
constructing a curve that represents the approximate envelope of noise
energy in the spectrum. This curve is computed from geometric median
and envelope followers constructed by smoothing the log spectrum with
a sliding window of approximately one Bark. The envelope curve is
companded according to the distance between the envelope and the mean;
greater distances imply greater tonality and lower distances imply
greater noisiness. Accordingly, the amplitude of this mask curve is
depressed in areas of greater tonality. Vorbis then adds a hardwired
bias curve (the \emph{noise offset}) to the companded envelope, producing
the final \emph{noise mask} curve.

\subsubsection{Noise Normalization}

Noise normalization is conceptually part of analyzing and handling
noise energy as described under noise masking, but it is handled in
a separate step. The purpose of normalization is to preserve approximate
wideband energy through quantization, especially at very low S/N ratios,
where large portions of the spectrum may otherwise collapse toward
silence. It may be viewed as a mechanism that preserves the intent
of noise masking through the quantization process. The need for the
specific implementation of noise normalization in the Vorbis encoder
is a consequence of the nature of a frequency domain codec.

\subsubsection{Impulse Analysis}

Impulse analysis refers to several metrics that characterize highly
temporally localized events in an audio frame, such as a sudden impulse
or attack in the audio. Earlier stages in the Vorbis encoder analyze
attacks and preecho potential to determine when to switch between
short and long frames, reusing these metrics in the frame-internal
coding algorithm. Frame analysis also inspects audio frames for an
impulse-train-like nature, that is, audio resembling a filter being
driven by an impulse train; such audio tends to have a particularly
regular harmonic structure into the high harmonics. Voice is the most
obvious example of this variety of audio.

As with noise normalization, this additional impulse analysis is necessitated
by the nature of a transform codec. Sudden impulses and the characteristic
tight \emph{rasp} of impulsive audio are features not compactly representable
in the frequency domain. Naive quantization causes narrow events to
smear in time and this loss of temporal resolution is perceptually
obvious. Impulse analysis is used to improve representation of non-sinusoidal,
non-random-noise content.

\subsubsection{Floor Construction}

The final floor curve in vorbis is created from the maximum of a direct
superposition of the tone mask and noise mask curves. The floor curve
is removed from the MDCT spectrum of a given audio frame, thus whitening
the spectrum and resulting in \emph{spectral residual} values. The
floor and residue are then coded via vector quantization.

\section{Application to CELP Coding}

\label{sec:Improved-Psycho-acoustics-in-CELP}

Speex does not use multiple blocksizes and for that reason, the block-switching
and analysis of the Vorbis codec is irrelevant. In addition, Speex
natively represents time-localized events and impulsive audio characteristics
very well. The impulse analysis as implemented in the Vorbis codec
is specific to transform codecs and as such is also not relevant to
Speex.

Similarly, noise normalization also addresses a need highly specific
to the frequency transform domain encoder. Noise normalization as
realized and used by the Vorbis encoder prevents gross wideband energy
inflation or collapse due to naive quantization, a situation from
which Speex is relatively well proofed by virtue of being an LPC-based
codec.

Tone and noise masking as described in Section \ref{sub:The-Vorbis-Model}
retain relevance in the context of the Speex codec. Experience with
the Vorbis encoder indicates that noise masking is responsible for
the greatest bulk of useful bit-rate savings. For purposes of initial
experimentation, we thus concentrate on implementing noise masking
alone in the Speex codec.

One of the assumptions made in the Vorbis codec is that the quantization
noise is entirely masked. After all, it must be for the codec to achieve
transparency or near-transparency. This assumption leads to using
a noise weighting curve that is very close to the masking curve, which
means a constant (negative) noise-to-mask ratio. However, the assumption
is not valid for Speex because there are simply not enough bits available
for the noise-to-mask ratio to be negative (or zero) at all frequencies.
This means that the quantization noise is always audible to some extent.
Using the masking curve directly for noise weighting in Speex would
results in over-emphasis of the noise in the high-energy regions of
the spectrum (typically low frequencies). For that reason, it is not
desirable to have a constant (positive) noise-to-mask ratio and the
masking curve used by Vorbis needs to be modified.

%A substantial difference between Vorbis and Speex is that Vorbis performs quantization of frequency (MDCT) domain data, while Speex quantizes in the time domain. Noise resulting from quantization in the frequency domain is narrowband by definition and undetectable until it produces audible preecho. Thus, Vorbis may use a noise weighting curve that is very close to the masking curve.
%Speex, however, performs quantization in the time domain, thus quantization noise is spread-spectrum. Using the masking curve directly for noise weighting in Speex would results in wideband leakage of the noise from the high-energy regions of the spectrum (typically low frequencies). For that reason, it is not desirable to have a constant (positive) noise-to-mask ratio and the masking curve used by Vorbis needs to be modified. 

In order to obtain good results with the Speex codec, we need to compress
the dynamic range of the masking curve. It was determined empirically
that the optimal companding consists of applying an exponent of $0.6$
to the masking curve computed by Vorbis. This value was found to be
suitable (near-optimal) for all bit-rates.

\begin{figure}
\begin{center}\includegraphics[width=1\columnwidth,keepaspectratio]{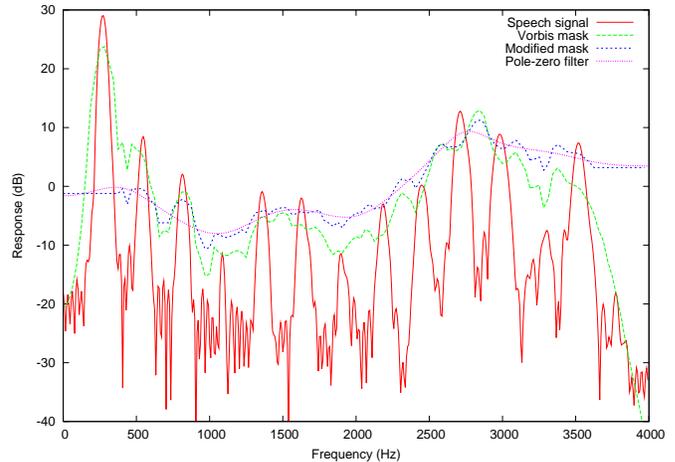}\end{center}

\caption{Modifications to the Vorbis mask. Arbitrary y-axis offset. \label{cap:Modifications-to-mask}}
\end{figure}

\subsection{Masking Curve to Weighting Filter}

In the integration of the new psychoacoustic model in Speex, it is
desirable to keep the same pole-zero formulation for the noise shaping
filter:\begin{equation}
\frac{1}{W(z)}=\frac{W_{n}(z)}{W_{d}(z)}\label{eq:weighting_num_den}\end{equation}

For that reason, the frequency-domain noise shaping curve is converted
to a $10^{th}$ order pole-zero filter using a two-step method. The
filter denominator $W_{d}(z)$ is first obtained by transforming the
masking curve into auto-correlation coefficients using an inverse
FFT and then applying the Levinson-Durbin algorithm. 

The filter numerator $W_{n}(z)$ is then estimated based on the error
between the all-pole model and the real masking curve. In practice,
$W_{n}(z)$ is nearly flat because $W_{d}(z)$ alone is able to provide
a very good approximation of the real curve.

Transforming the masking curve to a pole-zero model not only makes
the implementation easy, but it preserves the efficiency of the CELP
analysis-by-synthesis (AbS) codebook search. In order to limit the
complexity, the masking curve is computed only once every frame. For
each sub-frame, the curve is linearly interpolated and converted to
a pole-zero filter.

\subsection{Complexity Reduction\label{sub:Complexity-Reduction}}

The computation and conversion of the masking curve described above
tends to increases the complexity of the Speex codec. To compensate
for that, we propose three methods to minimize the impact of the proposed
psychoacoustic model:

\begin{enumerate}
\item Not using the filter numerator

Because the weighting filter numerator $W_{n}(z)$ generally has little
effect, it can be omitted without having a significant impact on quality.
This decreases the complexity of converting the masking curve to a
pole-zero filter, while simplifying the weighting filter. 

\item Setting the denominator to be the same as the synthesis filter

By forcing $W_{d}(z)=A(z)$, it is also possible to reduce the complexity.
While the complexity reduction in the conversion is smaller than in
method 1, the main advantage lies in the fact that the most commonly
used filter in the encoder (Fig. \ref{cap:Sub-frame-AbS}) simplifies
to $\frac{W(z)}{A(z)}=\frac{1}{W_{n}(z)}$. This reduces the complexity
of the encoder by approximately one million operations (add or multiply)
per second.

\item Using a frame-constant numerator

By using method 2 and keeping the same numerator $W_{n}(z)$ for whole
frames, we can reduce the cost of converting the filters by a factor
of four. The resulting complexity becomes similar to that of the reference
encoder, while still providing a quality improvement.

\end{enumerate}

\subsection{Application to Wideband Coding}

\begin{figure}
\includegraphics[width=1\columnwidth,keepaspectratio]{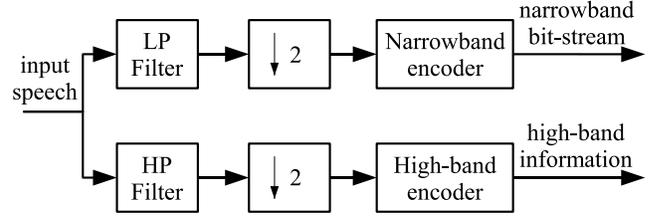}

\caption{Speex wideband encoder\label{cap:Speex-wideband-encoder}}
\end{figure}

To facilitate inter-operability with the public switched telephone
network (PSTN), Speex encodes wideband speech using the Sub-Band CELP
(SB-CELP) technique. This consists of splitting the signal in two
bands using a quadrature mirror filter (QMF), as shown in Fig. \ref{cap:Speex-wideband-encoder}.
The lower band (0-4 kHz) is encoded using the narrowband encoder,
while the higher band is encoded using a pitch-less version of CELP
(denoted HF-CELP). No pitch prediction is used for the higher band
because the spectral folding caused by the QMF makes the signal non-periodic. 

Only three parameters/vectors are transmitted: the high-band LSP parameters,
sub-frame gain corrections and the innovation signal. We use 8 LSPs,
jointly quantized with 12 bits. The gain corrections are computed
based on the ratio between high-band to low-band excitation and correcting
for the LPC response difference at the cutoff frequency (4 kHz). The
innovation can be encoded using 8, 4, or 0 kbps. In the lowest bit-rate,
only the shape of the spectrum is preserved and the excitation is
a frequency-aliased version of the narrowband part. This is done using
a technique conceptually similar to \cite{Valin2000} and requires
only 1.8 kbps to transmit the higher band.

Because of the embedded structure, no additional work is necessary
to make the proposed psychoacoustic model work with Speex in wideband
operation. It was found that the psychoacoustic model in the narrowband
encoder is also suitable for the lower band of wideband speech. For
the high-band, the reference psychoacoustic model is used. Although
it would be possible to compute the masking curve on the wideband
signal and then divide the spectrum in two bands, the added complexity
(code and CPU time) outweighs the potential benefits.

\section{Evaluation and Results}

\label{sec:Evaluation-and-Results}

We compare the reference Speex encoder to the modified encoder using
the Vorbis psychoacoustic model. The experiment is conducted using
Speex version 1.1.12, available from http://www.speex.org/ . The Vorbis
psychoacoustic model can be enabled by configuring with \texttt{-{}-enable-vorbis-psy}
or defining the \texttt{VORBIS\_PSYCHO} macro at compile time. 

\begin{figure}
\begin{center}\includegraphics[width=1\columnwidth,keepaspectratio]{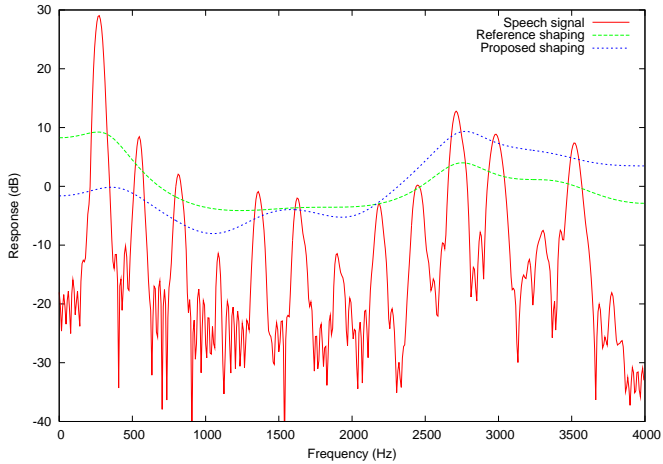}\end{center}

\caption{Noise shaping for reference encoder and modified encoder on a voiced
frame. Arbitrary y-axis offset.\label{cap:Noise-shaping-compar}}
\end{figure}

The Perceptual Evaluation of Speech Quality (PESQ) tool, as defined
by ITU-T recommendation G.862.2 \cite{Rix2001,P.862.2}, is used to
compare the encoders at various bit-rates for both narrowband and
wideband speech. While not a subjective mean opinion score (MOS) test,
we consider the results to be meaningful because we are only comparing
different noise-weighting filters for the same codec. The reference
Speex decoder is used for both encoders. The test set is composed
of 354 speech samples from 177 different speakers (87 male and 90
female) in 20 different languages taken from the NTT multi-lingual
speech database. The Speex codec reference implementation supports
variable search complexity. For the evaluation, the Speex variable
complexity option is set to 3, meaning that 3 simultaneous hypotheses
are updated when searching for the best adaptive and fixed codebook
entry.

\begin{figure}[t]
\begin{center}\includegraphics[width=1\columnwidth,keepaspectratio]{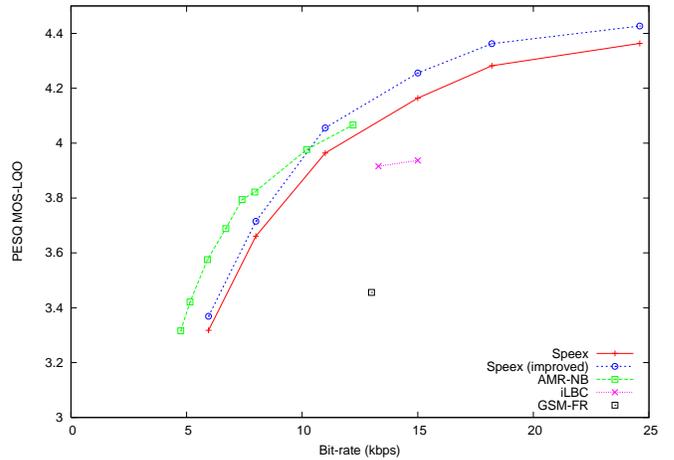}\end{center}

\caption{Quality of narrowband speech with and without improved noise model.\label{cap:Quality-of-narrowband}}
\end{figure}

Results for narrowband speech are presented in Fig. \ref{cap:Quality-of-narrowband}
with the AMR-NB, iLBC and GSM-FR codecs included as reference. Very
low Speex bit-rates (2.15 and 4 kbps) are not included because it
was found that the new model did not improve quality at those bit-rates.
It can be observed that the proposed noise weighting significantly
increases quality, especially at higher bit-rates. Also, it is worth
noting that the improved encoder at 11 kbps achieves the same level
of quality as the reference AMR-NB codec at 12.2 kbps.

\begin{figure}[t]
\begin{center}\includegraphics[width=1\columnwidth,keepaspectratio]{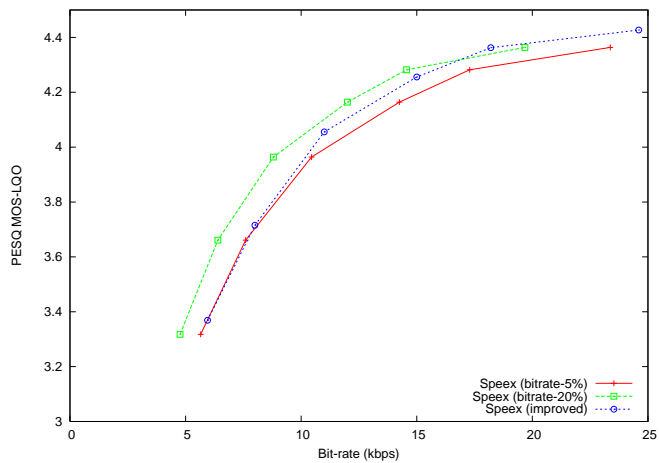}\end{center}

\caption{Bit-rate equivalent of the quality improvement.\label{cap:Bit-rate-equivalent}}
\end{figure}

In Fig. \ref{cap:Bit-rate-equivalent}, the quality of the original
and improved encoders are plotted with scaling of the x-axis by 5\%
and 20\% for the original encoder. It can be observed from there that
the improvement is equivalent to a bit-rate reduction of 5\% at low
bit-rates and up to 20\% at high bit-rates.

\begin{figure}
\begin{center}\includegraphics[width=1\columnwidth,keepaspectratio]{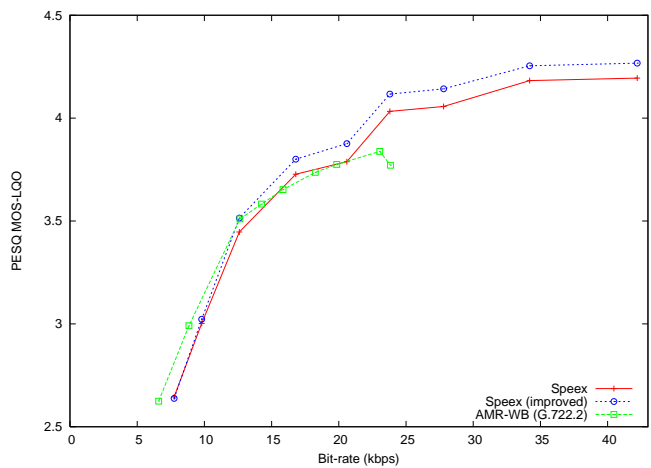}\end{center}

\caption{Quality of wideband speech with and without improved noise model.\label{cap:Quality-of-wideband}}
\end{figure}

Results for wideband speech are presented in Fig. \ref{cap:Quality-of-wideband}
with the AMR-WB codec included as reference. Again, very low Speex
bit-rates (3.95 and 5.8 kbps) are not included, since the quality
was not improved. The proposed noise weighting significantly increases
quality for wideband, clearly surpassing the quality of AMR-WB for
bit-rates above 12.8 kbps.

In a last experiment, we evaluate the effect of complexity reduction
methods proposed in Section \ref{sub:Complexity-Reduction}. Only
the 15 kbps narrowband mode is evaluated using each of the three complexity
reduction methods proposed. Results in Table \ref{cap:Impact-of-complexity}
show that methods C1 and C2 have no significant impact on quality,
while C3 only has a small negative impact. This makes C3 an attractive
choice, since it could lead to quality improvements without increasing
the encoder complexity.

\begin{table}
\begin{center}\begin{tabular}{|c|c|}
\hline 
Complexity&
PESQ MOS\tabularnewline
\hline
\hline 
proposed&
4.256\tabularnewline
\hline 
C1&
4.252\tabularnewline
\hline 
C2&
4.259\tabularnewline
\hline 
C3&
4.246\tabularnewline
\hline 
reference&
4.164\tabularnewline
\hline
\end{tabular}\end{center}

\caption{Impact of complexity reductions on speech quality. C1, C2 and C3
refer to the methods described in Section \ref{sub:Complexity-Reduction}.\label{cap:Impact-of-complexity}}
\end{table}

\section{Discussion and Conclusion}

\label{sec:Discussion-and-Conclusion}

In this work, we have demonstrated how to improve the quality of a
CELP codec by choosing a better psychoacoustic model. It has been
observed that the improvement is more significant at high bit-rate.
We find this result unintuitive because we expect it to be easier
to improve a lower quality modes. One hypothesis we propose to explain
this result is the fact that at higher bit-rates, there are more bits
available and thus more possibilities to change the bit allocation
based on the noise weighting. Another hypothesis is that the lower
the bit-rate, the further the optimal noise shaping is from the ideal
masking curve we use.

Unlike the work by \cite{Kamaruzzaman2004}, this proposed improvement
to Speex can be made without affecting compatibility or requiring
any modification on the decoder side. Also, because it only applies
to the encoder side, the technique could also improve the quality
of other existing and future CELP codecs. 

We believe this work clearly demonstrates that the noise weighting
currently used in CELP codecs has become inadequate. We have also
shown that improved noise weighting does not necessarily require increasing
the complexity of the encoder. For these reasons, it would be desirable
to further investigate alternative noise weighting filters for use
in CELP. We would also like to validate the results in this paper
with a subjective MOS test.

%Would need a real subjective test to confirm results.

\bibliographystyle{IEEEtran}
\bibliography{speex}

\begin{thebibliography}{10}
\providecommand{\url}[1]{#1}
\def\UrlFont{\rmfamily}
\providecommand{\newblock}{\relax}
\providecommand{\bibinfo}[2]{#2}
\providecommand\BIBentrySTDinterwordspacing{\spaceskip=0pt\relax}
\providecommand\BIBentryALTinterwordstretchfactor{4}
\providecommand\BIBentryALTinterwordspacing{\spaceskip=\fontdimen2\font plus
\BIBentryALTinterwordstretchfactor\fontdimen3\font minus
  \fontdimen4\font\relax}
\providecommand\BIBforeignlanguage[2]{{%
\expandafter\ifx\csname l@#1\endcsname\relax
\typeout{** WARNING: IEEEtran.bst: No hyphenation pattern has been}%
\typeout{** loaded for the language `#1'. Using the pattern for}%
\typeout{** the default language instead.}%
\else
\language=\csname l@#1\endcsname
\fi
#2}}

\bibitem{Schroeder1985}
M.~Schroeder and B.~Atal, ``Code-excited linear prediction({CELP}):
  High-quality speech at very low bit rates,'' in \emph{Proceedings IEEE
  International Conference on Acoustics, Speech and Signal Processing}, 1984,
  pp. 937--940.

\bibitem{Kroon1990}
P.~Kroon and B.~Atal, ``Pitch predictors with high temporal resolution,'' in
  \emph{Proceedings IEEE International Conference on Acoustics, Speech and
  Signal Processing}, 1990, pp. 661--664.

\bibitem{Chen1995}
J.-H. Chen, ``Toll-quality 16 kb/s {CELP} speech coding with very low
  complexity,'' in \emph{Proceedings IEEE International Conference on
  Acoustics, Speech and Signal Processing}, 1995, pp. 9--12.

\bibitem{Johnston1988}
J.~Johnston, ``Transform coding of audio signals using perceptual noise
  criteria,'' \emph{IEEE Journal of Selected Areas in Communication}, vol.~6,
  pp. 314--323, 1988.

\bibitem{Lefebvre1994}
R.~Lefebvre, R.~Salami, C.~Laflamme, and J.-P. Adoul, ``High quality coding of
  wideband audio signals using transform-coded excitation ({TCX}),'' in
  \emph{Proceedings IEEE International Conference on Acoustics, Speech, and
  Signal Processing}, vol.~I, 1994, pp. 193--196.

\bibitem{Ehmer1959}
R.~Ehmer, ``Masking patterns of tones,'' \emph{Journal of the Acoustical
  Society of America}, vol.~31, no.~8, pp. 1115--1120, 1959.

\bibitem{Valin2000}
J.-M. Valin and R.~Lefebvre, ``Bandwidth extension of narrowband speech for low
  bit-rate wideband coding,'' in \emph{Proceedings IEEE Speech Coding
  Workshop}, 2000, pp. 130--132.

\bibitem{Rix2001}
A.~Rix, J.~Beerends, M.~Hollier, and A.~Hekstra, ``Perceptual evaluation of
  speech quality ({PESQ}) -- a new method for speech quality assessment of
  telephone networks and codecs,'' in \emph{Proceedings IEEE International
  Conference on Acoustics, Speech, and Signal Processing}, 2001.

\bibitem{P.862.2}
ITU-T, \emph{Recommendation P.862.2 -- Wideband extension to Recommendation
  P.862 for the assessment of wideband telephone networks and speech codecs},
  International Telecommunications Union, 2005.

\bibitem{Kamaruzzaman2004}
M.~Kamaruzzaman and H.~Taddei, ``Embedded speech codec based on speex,'' in
  \emph{Proceedings of the 116th AES Convention}, 2004.

\end{thebibliography}

\end{document}